\documentclass{article}

\usepackage{hyperref}
\usepackage{amsmath, amsthm, amssymb}
\usepackage{graphicx}

\newtheorem{thm}{Theorem}

\newtheorem*{rem}{Remark}

\begin{document}

\title{On a~mechanical lens}
\author{A.P.~Ivanov$^1$, N.N.~Erdakova$^2$\vspace{3mm}\\
{\small $^1$Moscow Institute of Physics and Technology,}\\ {\small Institutskii per., 9, Dolgoprudny, Moscow Region,141700, Russia}\\
{\small $^2$Udmurt State University,}\\ {\small Universitetskaya st. 1,  Izhevsk, 426034, Russia}}

\maketitle

\begin{abstract}

In this paper, we consider the dynamics of a heavy homogeneous ball moving under the influence of dry friction on a fixed horizontal plane.
We assume the ball to slide without rolling. We demonstrate that the plane may be divided into two regions, each characterized by a distinct
coefficient of friction, so that balls with equal initial linear and angular velocity will converge upon the same point from different
initial locations along a certain segment. We construct the boundary between the two regions explicitly and discuss possible applications
to real physical systems.
\end{abstract}

\section{Introduction}\label{sec1}



Systems with friction continue to be an area of intense interest.
It is well known that friction is of fundamental importance in problems involving sports dynamics such as billiards,
bowling, curling, motion of the skateboard,
and others.
Unfortunately, the problems of dynamical friction which occur during the motion are poorly understood.
The few studies worth mentioning are concerned with the stability of decelerative sliding motions of a driven mechanical system~\cite{Vielsack},
the motion of a cylinder on a rough plane~\cite{BEIM}, and the motion of a curling rock~\cite{Shagelski} and~\cite{Ivanov}.

The effects of friction are usually described using the
nonholonomic model. This model is a simplification of the initial
systems with friction: the coefficient of friction tends to infinity
and the motion is assumed to occur on an absolutely rough plane. The classical results on nonholonomic dynamics which go back to the
work of Routh, Appell, Chaplygin, Zhukovskii, and others (see for ex.~\cite{Chaplygin,
zhuk}) are well known.
Some recent results on the dynamics of nonholonomic systems can be found in the paper by Batista~\cite{Batista1, Batista2}, where the motion of disks on a
plane is studied, and in the paper~\cite{BMK2}, where the motion of a ball is considered.
It should be noted that the behavior of such nonholonomic
systems exhibits strange, unusual dynamics. The problem with especially demonstrative behavior in this
sense is the motion of Celtic stone~\cite{BMK4}.

\begin{figure} [!ht]
\centering
\includegraphics{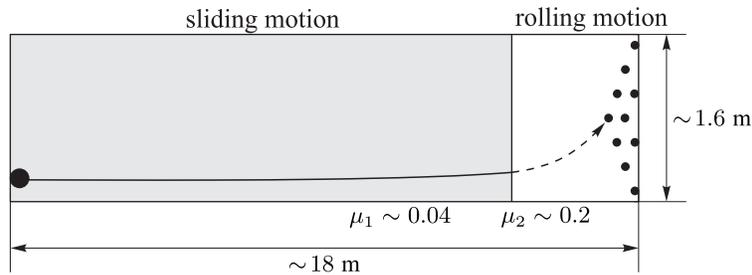}
\caption{Scheme of bowling. Oiled path of sliding (the solid line) and rough path of rolling (the dashed line) of a~ball.}
\label{fig001}
\end{figure}

It is well known that sliding and rolling phases can alternate in the problem of
motion of a ball on a plane. The most famous and popular dynamical game based on the dynamical properties of the ball is
bowling.
 It seems that a~ball
thrown from the hands of a~professional works miracles. However,
theoretical and applied studies carried out over the last fifty years have
shown that, on the one hand, under the simplest initial conditions and
parameters bowling is quite a~determined game, but, on the other hand,
there are a~lot of unexplained effects observed in professional bowling
which still remain unexplained. The motion of a~ball in professional
bowling can be divided into two phases. The first one is a~slightly curled
(the ball moves in a~parabola) sliding of the ball on a~fairly smooth
oiled surface with coefficient of friction about $\sim 0.04$. The second
one is the passage of the ball to a~dry surface (the coefficient of
friction is $\sim0.2$) with subsequent rolling without slipping and a
remarkable phenomenon of hook~--- a~sharp curl of the trajectory during the
terminal motion (fig.~\ref{fig001}).

While the first stage is a~well known effect of motion of a~ball in a
parabola during sliding, studied by Euler~\cite{1}, the
second stage is a~more complex motion studied in a~large body of
literature. It is well known that a~homogeneous ball rolls in a~straight
line~\cite{Chaplygin}, this fact is also illustrated with
bowling~\cite{hopkins}. But this effect can be observed in unprofessional
amateur bowling, where paths and balls are not prepared in a~special way.
A great deal of classical research is devoted to the dynamics of a~ball
with nonuniform mass distribution~\cite{zhuk, BMK2}, where it is shown
that the trajectory of a~ball deviates from a~straight line during the
rolling motion. Some papers are directly concerned with explanation and
prediction of the dynamics of special professional bowling
balls~\cite{frohlich,King} with emphasis on the final instant of the
ball's motion~--- hook.

\begin{figure} [!ht]
\centering
\includegraphics{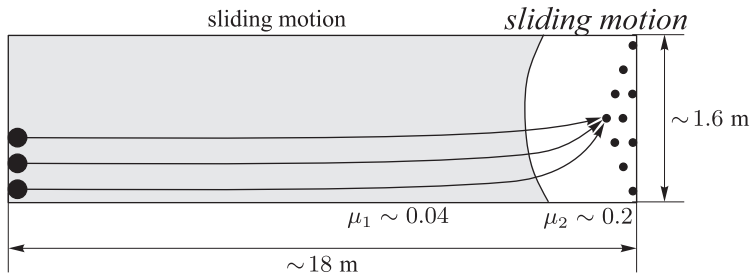}
\caption{Scheme of a~``mechanical lens'' in bowling. The balls slide from
the starting points till the target point without rolling.} \label{fig01}
\end{figure}

However, as stated above, the ball starts a~curl at the stage of sliding.
This curling depends on the coefficient of friction. The question arises
whether we can can we reach the effect of hook on the stage of sliding
before the transition to rolling, for example, when the ball passes from
the smoother to the rougher surface of the path. It is also of interest to
consider a~more complicated problem of a~sliding ball. Assuming the
coefficient of friction to be variable, we could calculate the boundary
between the surfaces in the path in such a~way that parallel families of
trajectories of sliding analogous balls with equal initial conditions
(linear and angular velocities) converges to a~predetermined point, for
example, to the central skittle in the bowling\footnote{In particular this
problem was discussed by the authors of this paper and professor Andy
Ruina in the course of the IUTAM Symposium http://iutam2012.rcd.ru/}
(fig.~\ref{fig01}). We call this phenomenon~--- the effect of ``mechanical
lens'' by analogy with the optical effect of the well known collecting
lens focusing the light beams in  one point.

Thus, this paper is devoted to analytical and numerical studies of the
effect of the curling of a~trajectory and the effect of a~``mechanical
lens'' in the dynamics of a~ball during sliding. Also, we study a~possible
application of these phenomena to the bowling game.

\section{Equations of motion of a~sliding ball}\label{sec2}

Consider a~heavy homogeneous ball moving by inertia on a~fixed rough
horizontal plane. We assume the velocity of the point of contact to be
sufficiently large to neglect the spinning friction and rolling friction
and their influence on the law of sliding friction. For the latter we take
the Coulomb formula
\begin{equation}
\label{eq1}
\boldsymbol{F}=-\mu P\frac{\boldsymbol{u}}{u},
\end{equation}
where $\boldsymbol{F}$ is the friction force, $P$ is the weight of the ball, $\mu$
is the coefficient of friction and $\boldsymbol{u}$ is the velocity of the point
of contact. This system was first investigated in 1758 by Johann Euler
(a~son of Leonhard Euler)~\cite{1}. We list the basic properties of motion which are
important in what follows.
\begin{enumerate}
  \item $\boldsymbol{e}=\frac{\boldsymbol{u}}{u}=\mbox{const}$, i.e. the direction of sliding does not change;
  \item if the initial velocity of the center of the ball $\boldsymbol{v}_0 $ is not collinear to
$\boldsymbol{e}=\frac
{\boldsymbol{u}}{u}$, then the center of the ball moves in a~parabola (until the ball stops sliding)
\begin{equation}
\label{eq2}
\boldsymbol{r}=\boldsymbol{r}_0 +\boldsymbol{v}_0 t+\frac{1}{2}ft^2\boldsymbol{e},\quad f=\mu g,
\end{equation}
where $\boldsymbol{r}$ is the radius vector of the point of contact, $\boldsymbol{r}_0
$ is its initial value (for $t=0)$, $t$ is the duration of the motion
and $g$ is the free-fall acceleration;
   \item The absolute value of the sliding velocity decreases by the law
\begin{equation}
\label{eq3}
u=u_0 -f\left( {1+\frac{R^2}{\rho ^2}} \right)t,
\end{equation}
where $u_0 $ is the initial value and $R$ and $\rho $ are the radius of the ball and its radius of inertia, respectively.
\end{enumerate}

These properties allow the trajectory of the center of the ball to be uniquely constructed.

\begin{figure} [!ht]
\centering
\includegraphics[scale=0.6]{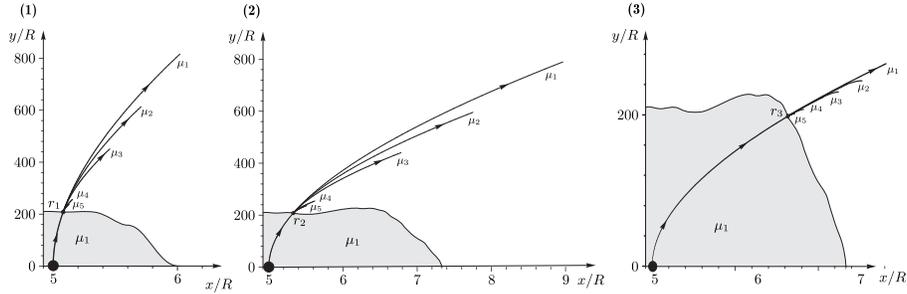}
\caption{Deviation of three families of trajectories of a~sliding bowling ball from a~straight line during the sliding motion
 when the coefficient of friction changes from $\mu_1=0.04$
to different values: $\mu_2=0.06$, $\mu_3=0.08$, $\mu_4=0.1$, $\mu_5=0.5$. The
mass and radius of a~ball are $m=5$~kg, $R=0.1$~m, respectively. For the
first family \textbf{(1) }of trajectories the initial linear and angular velocities are
$\boldsymbol{v_0}=[0;10]$~m/s, $\boldsymbol{w_0}=[0;1]$~s$^{-1}$, for the second \textbf{(2) }
$\boldsymbol{v_0}=[0;10]$~m/s, $\boldsymbol{w_0}=[3;4]$~s$^{-1}$, for the third \textbf{(3) }
$\boldsymbol{v_0}=[0;6]$~m/s, $\boldsymbol{w_0}=[5;5]$~s$^{-1}$.} \label{fig02}
\end{figure}

Using equations~\eqref{eq2} and~\eqref{eq3} it is not difficult to show
qualitative and quantitative changes of trajectories of a~sliding ball
during the passage from the smoother to the rougher surface of the path.
In Figure~\ref{fig02} the families of trajectories of a~sliding bowling
ball with different initial linear and angular velocities are shown. At
the points $r_i$ the coefficient of friction changes from $\mu_1=0.04$ to
$\mu_2=0.06$, $\mu_3=0.8$, $\mu_4=0.1$, $\mu_5=0.5$. As evident from
Fig.~\ref{fig02} the growth of the coefficient of friction leads to an
earlier termination of the sliding motion of the ball and to a~more
significant curling of its trajectory. It is also clear that both
quantities strongly depend on initial conditions of the system.

Assuming $u=0$  in~(\ref{eq3}), it not difficult to define the time of
sliding of the ball on the surface with the coefficient of friction $\mu$
$$
t=\frac{2u_0}{7\mu g}
$$
and to estimate it for different couples of surfaces. For example, for the
family $\textbf{(1)}$ the total time of sliding is $t=2.41$~sec for the
couple $(\mu_1;\mu_5)$, whereas if the ball slides on a~homogeneous
surface with $\mu_1$, the total time is $t=7.14$~sec.

\section{Equation of the boundary curve between two surfaces}\label{sec3}

Now assume that the coefficient of friction is variable. Let us calculate
the boundary between the surfaces on the path in such a~way that parallel
families of analogous homogenous sliding balls launched under equal
initial conditions (linear and angular velocities) on a~horizontal rough
plane converge to a~predetermined point. Let us write the analytical
equation for the curve that is the  boundary between the surfaces.

\begin{figure} [!ht]
\centering
\includegraphics[scale=0.9]{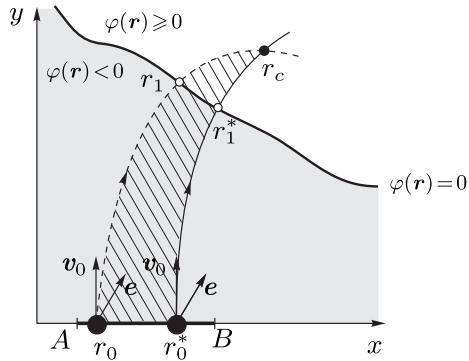}
\caption{Scheme of the model of ``mechanical lens''. The curve $\varphi
\left( {\boldsymbol{r}} \right)=0$ corresponds to the boundary curve
between two surfaces. The shaded area is a~domain of attraction of the
system.} \label{fig003}
\end{figure}

Let us choose a~starting segment $[AB]$ (without loss of generality we set
$[AB]\in Ox$) and assume that for all trajectories emanating from it the
vectors $\boldsymbol{v}_0 $ and $\boldsymbol{e}$ are equal and
noncollinear (see Fig.~\ref{fig003}). According to formula (\ref{eq2}), at
$t>0$ this segment moves uniformly. We take the dependence of the
coefficient of friction on $\boldsymbol{r}$ to be binary, i.e.
\begin{equation}
\label{eq4}
f\left( {\boldsymbol{r}} \right)=
\left\{
\begin{aligned}
&f_1 , &&\mbox{if}\quad \varphi \left( {\boldsymbol{r}} \right)<0 \\
&f_2 , &&\mbox{if}\quad \varphi \left( {\boldsymbol{r}} \right)\ge 0
\end{aligned} \right.
\end{equation}
where $f_1 \ne f_2 $ and the function $\varphi \left( {\boldsymbol{r}}
\right)$ specifying the boundary curve between two surfaces for $\varphi
\left( {\boldsymbol{r}} \right)=0$ must be defined.
Let us take a~point $\boldsymbol{r}_0^\ast $ from the interval $(AB)$ as
the initial value and construct the trajectory \eqref{eq2} emanating from
it. We shall assume this trajectory to be supporting and choose on it a
target point $C$ which must be reached by all trajectories sufficiently
close to this one. To do this, we fix two quantities: $t_1 >0$ is the time
of motion in the region $\varphi \left( {\boldsymbol{r}} \right)<0$ and
$s>0$ is the time of motion in the region $\varphi \left( {\boldsymbol{r}}
\right)>0$. Such a~choice is sufficiently arbitrary, we only need to make
sure (formula \eqref{eq3}) that the friction has no time to stop the
sliding motion of the ball. In addition, the quantity $t_1$ should not be
small, since otherwise the boundary curve between two surfaces can cross
the starting segment, which will narrow the family of balls.  Due to
\eqref{eq2} and \eqref{eq4} the equation of the supporting trajectory is
\begin{equation}
\label{eq5}
\begin{gathered}
 \boldsymbol{r}^\ast (t)=\left\{
\begin{aligned}
&\boldsymbol{r}_0^\ast +\boldsymbol{v}_0 t+\frac{1}{2}f_1 t^2\boldsymbol{e},&&\text{if}\quad t\le t_1 \\
&\boldsymbol{r}_1^\ast +\boldsymbol{v}_1^\ast (t-t_1 )+\frac{1}{2}f_2 (t-t_1 )^2\boldsymbol
{e},&&\mbox{if}\quad t_1 \le t\le t_1 +s\\
\end{aligned} \right. \\
\boldsymbol{r}_1^\ast =\boldsymbol{r}_0^\ast +\boldsymbol{v}_0 t_1
+\frac{1}{2}f_1 t_1^2 \boldsymbol{e},\quad \boldsymbol{v}_1^\ast
=\boldsymbol{v}_0 +f_1 t_1 \boldsymbol{e}.
\end{gathered}
\end{equation}
The radius vector of the target point $\boldsymbol{r}_C $ is defined from the formula
\begin{equation}
\label{eq6}
\boldsymbol{r}_C =\boldsymbol{r}_1^\ast +\left(
{\boldsymbol{v}_0 +f_1 t_1 \boldsymbol{e}} \right)s+\frac{1}{2}f_2
s^2\boldsymbol{e},
\end{equation}
and the boundary curve between two surfaces passes through the point
$\boldsymbol{r}_1^\ast $, i.e. $\varphi \left( {\boldsymbol{r}_1^\ast }
\right)=0$.

We now consider another trajectory from this family. Its initial point
$\boldsymbol{r}_0 $ in the interval $(AB)$ is defined by the number
$\delta $ (proportional to the distance between $\boldsymbol{r}_0^\ast $
and $\boldsymbol{r}_0 )$, so that
\begin{equation}
\label{eq7}
\boldsymbol{r}_0 =\boldsymbol{r}_0^\ast +\delta \boldsymbol{i}=\boldsymbol{r}_0^\ast +\delta (\alpha
\boldsymbol{v}_0 +\beta \boldsymbol{e}),
\end{equation}
where $\alpha ,\beta $ are the coordinates of the direction vector $\boldsymbol{i}$ of the straight line
$(AB)$ in the skew-angular basis $(\boldsymbol{v}_0 ,\;\boldsymbol{e})$. By analogy with (\ref{eq5}), the equation of this trajectory is
\begin{equation}
\label{eq8}
\begin{gathered}
 \boldsymbol{r}(t)=\left\{
\begin{aligned}
&\boldsymbol{r}_0 +\boldsymbol{v}_0 t+\frac{1}{2}f_1 t^2\boldsymbol{e},&&\mbox{if }
t\le t_1 +\tau \\
&\boldsymbol{r}_1 +\boldsymbol{v}_1 (t-t_1 -\tau )+\frac{1}{2}f_2 (t-t_1 -\tau
)^2\boldsymbol{e},&&\mbox{if } t_1 +\tau \le t\le t_1 +\tau +p\\
\end{aligned} \right. \\
\boldsymbol{r}_1 =\boldsymbol{r}_0 +\boldsymbol{v}_0 (t_1 +\tau
)+\frac{1}{2}f_1 (t_1 +\tau )^2\boldsymbol{e},\quad \boldsymbol{v}_1 =\boldsymbol{v}_0 +f_1
(t_1 +\tau )\boldsymbol{e},
 \end{gathered}
\end{equation}
where $\boldsymbol{r}_0 $ is defined by \eqref{eq7}, the instant of time
$t_1 +\tau $ corresponds to the intersection of the trajectory with the
boundary curve between two surfaces, and the quantity $p$ is equal to the
time of motion of the ball in the region $\varphi >0$.

The condition for the trajectory to reach the point $C$ is expressed by the equality
\begin{equation}
\label{eq9}
\boldsymbol{r}(t_1 +\tau +p)=\boldsymbol{r}_C
\end{equation}
The vector equality \eqref{eq9} is equivalent to the system of two scalar
equations in two unknowns, $\tau $ and $p$, which also contains the
parameter $\delta $ defining the initial position of the trajectory from
the family. Hence, one can express two of these quantities as some
functions of the third one (e.g., $p$ and $\delta $ in terms of $\tau )$.
Then the formula
\begin{equation}
\label{eq10}
\boldsymbol{r}=\boldsymbol{r}_1 =\boldsymbol{r}_1^\ast +\delta (\alpha \boldsymbol{v}_0 +\beta \boldsymbol
{e})+\boldsymbol{v}_0 \tau +\frac{1}{2}f_1 (2t_1 \tau +\tau ^2)\boldsymbol{e}
\end{equation}
is a~parametric equation of the boundary curve between two surfaces. In
particular, if $\tau =0$ and $\delta =0$, we obtain
$\boldsymbol{r}=\boldsymbol{r}_1^\ast$, which corresponds to the switching
point on the supporting curve~\eqref{eq5}. Consequently, the formula
\eqref{eq10} solves the problem.

All trajectories starting from the $\delta$-neighborhood~of
$\boldsymbol{r}_0^\ast$ with initial conditions $(\boldsymbol{v}_0
,\;\boldsymbol{e})$ and reaching the target point $\boldsymbol{r}_C$
generate the \textit{domain of attraction} of the system (see
Fig.~\ref{fig003}).

\begin{rem}
After constructing the boundary curve between two surfaces \eqref{eq10},
we have to make sure that the supporting trajectory \eqref{eq5} crosses
it: in some degenerate cases, touching with return into the region
$\varphi <0$ is possible. In addition, it is necessary to make sure that
there are no repeated intersections with these lines.
\end{rem}

Substituting \eqref{eq6}, \eqref{eq7} and \eqref{eq10} in \eqref{eq9}, we obtain
\begin{equation}
\label{eq11}
\begin{gathered}
 \delta (\alpha \boldsymbol{v}_0 +\beta \boldsymbol{e})+\boldsymbol{v}_0 \tau +\frac{1}{2}f_1
(2t_1 \tau +\tau ^2)\boldsymbol{e}+
 +\left( {\boldsymbol{v}_0 +f_1 \left( {t_1 +\tau } \right)\boldsymbol{e}}
\right)p+\frac{1}{2}f_2 p^2\boldsymbol{e}=\\
=\left( {\boldsymbol{v}_0 +f_1 t_1 \boldsymbol{e}}\right)s+\frac{1}{2}f_2 s^2\boldsymbol{e}.
 \end{gathered}
\end{equation}
Equating the coefficients for the basis vectors $\boldsymbol{v}_0 $ and
$\boldsymbol{e}$ on the left-hand and right-hand sides of (\ref{eq11}), we
obtain the system
\begin{gather}
\label{eq12}
\alpha \delta +\tau +p=s,
\\
\label{eq13}
\beta \delta +f_1 t_1 \tau +\frac{1}{2}f_1 \tau ^2+f_1 (t_1 +\tau
)p+\frac{1}{2}f_2 p^2=f_1 t_1 s+\frac{1}{2}f_2 s^2.
\end{gather}
Eq. \eqref{eq12} is linear and allows us to eliminate one of the variables without difficulty. Eq. (\ref{eq13}) is quadratic, and it takes extra effort to use it.

\section{The case of an absolutely smooth surface on one of the phases of motion}\label{sec4}

We point out two limiting particular cases where Eq.~\eqref{eq13} simplifies.
\begin{itemize}
  \item $\boldsymbol{f_1 =0}$, $\boldsymbol{f_2 \ne 0}$, i.е. the ball
      moves first on the smooth part of the plane (``ice'') and then
      gets onto the rough part.

Eq. \eqref{eq13} becomes
\[
\beta \delta +\frac{1}{2}f_2 p^2=\frac{1}{2}f_2 s^2.
\]
If $\beta \ne 0$ (i.e. the starting segment $[AB]$ is
not collinear to the initial velocity $\boldsymbol{v}_0 )$, then
\[
\delta =\frac{1}{2}\beta ^{-1}f_2 \left( {s^2-p^2} \right),\quad \tau
=s-\alpha \delta -p.
\]
In~\eqref{eq10} we obtain a~parametric equation of the boundary curve
between two surfaces in the form
\begin{equation}
\label{eq14}
\boldsymbol{r}(p)=\boldsymbol{r}_1^\ast +\delta (\alpha \boldsymbol{v}_0 +\beta \boldsymbol{e})+\tau
\boldsymbol{v}_0 =\boldsymbol{r}_1^\ast +\frac{1}{2}f_2 \left( {s^2-p^2} \right)\boldsymbol
{e}+(s-p)\boldsymbol{v}_0,
\end{equation}
where $\delta =0$ and $\boldsymbol{r}=\boldsymbol {r}_1^\ast $
correspond to the value of the parameter $p=s$, i.e. we are on the
supporting curve. Eq.~\eqref{eq14} defines the curve of order~2, which
is obviously a~parabola, since it is unbounded and connected.

In the case $\beta =0$ all trajectories emanating from the segment
$\left[ {AB} \right]$ merge to form a~single (supporting) trajectory,
although the switching point $\boldsymbol {r}_1^\ast $ is reached at
different instants of time.

\item $\boldsymbol{f_1 \ne 0,\;f_2 =0}$, i.e. the ball moves first on
    the rough part of the plane and then gets onto ``ice''.

Eq.~\eqref{eq13} becomes
\[
\beta \delta +f_1 t_1 \tau +\frac{1}{2}f_1 \tau ^2+f_1 (t_1 +\tau )p=f_1 t_1
s.
\]
Substituting~\eqref{eq12} into the above equation gives
\begin{equation}
\label{eq15}
\delta \left( {\beta -\alpha f_1 (t_1 +\tau )} \right)-\frac{1}{2}f_1 \tau
^2+f_1 \tau s=0.
\end{equation}
In this case it is more convenient to use $\tau $ as a~parameter. If
\begin{equation}
\label{eq16}
\beta -\alpha f_1 t_1 \ne 0,
\end{equation}
then the coefficient with $\delta $ in~\eqref{eq15} is different from
zero in a~sufficiently small neighborhood of the value $\tau =0$. Then
\begin{equation}
\label{eq17}
\delta =f_1 \tau \left( {\frac{1}{2}\tau -s} \right)\left( {\beta -\alpha
f_1 (t_1 +\tau )} \right)^{-1}.
\end{equation}
Substituting~\eqref{eq17} in~\eqref{eq10}, we obtain a~representation
of the boundary curve between two surfaces in the form of a~rational
parametric curve
\[
\boldsymbol{r}(\tau )=\boldsymbol{r}_1^\ast +f_1 \tau \left( {\frac{1}{2}\tau -s}
\right)\left( {\beta -\alpha f_1 (t_1 +\tau )} \right)^{-1}(\alpha \boldsymbol
{v}_0 +\beta \boldsymbol{e})+\boldsymbol{v}_0 \tau +\frac{1}{2}f_1 (2t_1 \tau +\tau
^2)\boldsymbol{e},
\]
Note that the value $\tau =0$ corresponds to the supporting trajectory.

In the case where an equality takes place in formula~\eqref{eq16}, \eqref{eq15}
becomes
\[
f_1 \tau \left( {\delta \alpha -\frac{1}{2}\tau +s} \right)=0.
\]
This means that either $\tau =0$ and $\delta $ is arbitrary, or
$\delta \alpha -\frac{1}{2}\tau +s=0$. In the former case,
formula~\eqref{eq10} describes a~segment parallel to the starting
segment $[AB]$. However, it turns out that when the boundary curve
between two surfaces is reached the trajectory touches this curve, and
then all trajectories pass along the switching segment, the target
point $C$ also lies on this segment.

In the latter case, $\alpha \ne 0$, otherwise by virtue of the
equality opposite to~\eqref{eq16}, we would also have $\beta =0$,
which is impossible, since $\boldsymbol{i}\ne 0$. Hence, $\delta $
will be a~linear function of $\tau $, and the formula~\eqref{eq10}
describes a~parabola (by analogy with the case $1^\circ$). However,
such a~curve does not contain the point of the supporting trajectory
$\tau =0$, $\delta =0$. Therefore, it cannot be regarded as a~solution
to the problem.
\end{itemize}

\section{Analysis of motion in the general case}\label{sec5}

We now turn to a~discussion of the general case $0<f_1 \ne f_2 >0$. By
using~\eqref{eq12} twice, we bring~\eqref{eq13} to the form
\[
\delta (\beta -f_1 t_1 )+\frac{1}{2}f_1 (\alpha ^2\delta ^2-2\alpha \delta
s)+\frac{1}{2}\left( {f_2 -f_1 } \right)p^2=\frac{1}{2}\left( {f_2 -f_1 }
\right)s^2,
\]
whence
\begin{equation}
\label{eq18}
p(\delta )=\left[ {s^2-\left( {2\delta (\beta -\alpha f_1 t_1 )+f_1 (\alpha
^2\delta
 ^2-2\alpha \delta s)} \right)\left( {f_2 -f_1 } \right)^{-1}}
\right]^{1/2}.
\end{equation}
The sought-for curve is irrational. It is governed by the formula
\begin{equation}
\label{eq19}
\boldsymbol{r}(\delta )=\boldsymbol{r}_1^\ast +\delta (\alpha \boldsymbol{v}_0 +\beta \boldsymbol
{e})+\boldsymbol{v}_0 \tau +\frac{1}{2}f_1 (2t_1 \tau +\tau ^2)\boldsymbol{e},
\end{equation}
\[
\tau =s-p(\delta )-\alpha \delta,
\]
where $p(\delta )$ is expressed by~\eqref{eq18}. We note that for $\delta
=0$ we have $p=s$ and $\tau =0$, which corresponds to the supporting
trajectory.

We make sure that the curve~\eqref{eq19} intersects the supporting
trajectory, i.e. the vector $\boldsymbol {v}_1^\ast =\boldsymbol{v}_0 +f_1
t_1 \boldsymbol{e}$ is not tangential to this curve. The tangent vector to
the curve (\ref{eq19}) at the point of its intersection with the
supporting trajectory is defined by the formula
\begin{equation}
\label{eq20}
\left. {\frac{d\,\boldsymbol{r}(\delta )}{d\,\tau }} \right|_{\tau =0} =\left.
{\frac{d\,\delta }{d\,\tau }} \right|_{\tau =0} (\alpha \boldsymbol{v}_0 +\beta
\boldsymbol{e})+\boldsymbol{v}_0 +f_1 t_1 \boldsymbol{e}.
\end{equation}
It follows from~\eqref{eq18} and~\eqref{eq19} that the derivative on the
right-hand side of~\eqref{eq20} is different from zero under the condition
$p\ne 0$, which is obviously satisfied in a~neighborhood of the supporting
trajectory (on which $p=s)$. Consequently, the no-touching condition is
equivalent to the non-collinearity of the vectors $\boldsymbol{v}_1^\ast $
and $\boldsymbol{i}=\alpha \boldsymbol{v}_0 +\beta \boldsymbol{e}$, i.e.
\begin{equation}
\label{eq21}
\beta \ne \alpha f_1 t_1.
\end{equation}
We note that the condition~\eqref{eq21} is equivalent to~\eqref{eq16}.

Summarizing the investigation, we formulate the main result in the form of
a~theorem.

\begin{thm}
Suppose there is a~family of balls which is characterized by the initial
velocity of the center $\boldsymbol{v}_0 $ and by the unit vector opposing
the sliding velocity of the point of contact $\boldsymbol{e}$, and
positive coefficients of friction $\mu _1 \ne \mu _2 $. The initial
positions of the points of contact lie on the segment of a~straight line
with the direction vector $\boldsymbol{i}=\alpha \boldsymbol{v}_0 +\beta
\boldsymbol{e}$. Let one of the trajectories emanating from this segment
be the supporting trajectory. The motion along the supporting trajectory
consists of two phases (see~\eqref{eq5}). The duration of the first phase
$t_1 $ is determined according to~\eqref{eq21}. The target point $C$ lies
on the supporting trajectory and is determined by the duration of the
second phase $s$ whose value is limited by the sliding condition (in the
formula~\eqref{eq3} $u>0)$.

Then there exists a~unique boundary curve between two surfaces during the
intersection with which the coefficient of friction changes from $\mu _1 $
to $\mu _2 $, such that for sufficiently small $\vert \delta \vert $ all
trajectories of the family pass through the point $C$.
\end{thm}

An analogous assertion holds in the limiting cases $\mu _1 =0$ (under the
condition $\beta \ne 0)$ and $\mu _2 =0$. If $\mu _1 =0,\,\beta =0$, then
all trajectories merge into a~single one (with time shift), and the
boundary curve between two surfaces is not determined.

\section{Examples of focusing trajectories of a~sliding balls}

Using the above algorithm, we construct the curves of the boundary between
the surfaces~\eqref{eq19}, supporting trajectories~\eqref{eq6} of a
sliding ball and trajectories from its $\delta$-neighborhood~\eqref{eq8}
for different couples of surfaces: for the cases of a~passage to the
smoother or to the rougher plane.

\begin{figure} [!ht]
\centering
\includegraphics[scale=0.9]{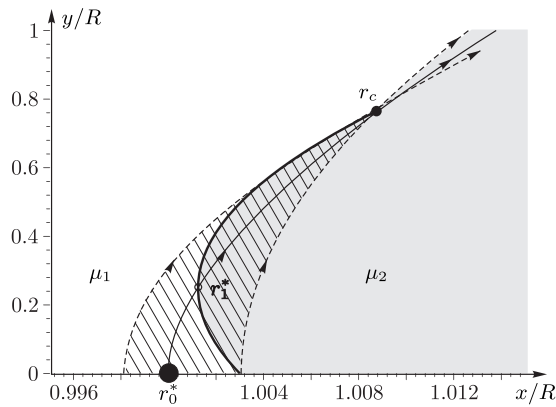}
\caption{Supporting trajectory (thin solid line), the curve of the
boundary between two surfaces (thick solid line), the point of boundary
between two surfaces on the supporting trajectory $\boldsymbol{r}_1^*$,
the target point $\boldsymbol{r}_С$ and a~couple of trajectories from a
family (dashed lines) for the case $\mu _1 =0.2$, $\mu _2 =0.1$. The shaded
area is a~domain of attraction of the system.} \label{fig1}
\end{figure}

\begin{figure} [!ht]
\centering
\includegraphics[scale=0.9]{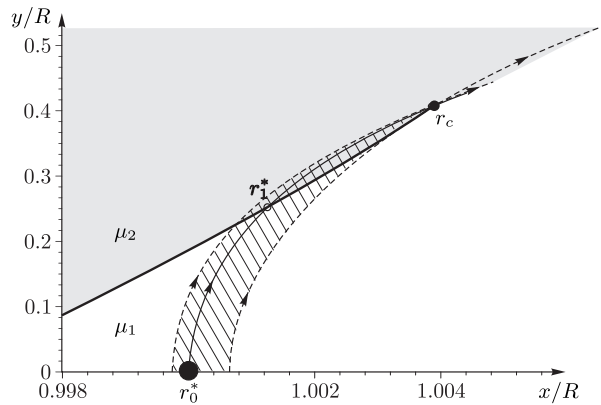}
\caption{Supporting trajectory (thin solid line), the curve of boundary
between two surfaces (thick solid line), the point of boundary between two
surfaces on the supporting trajectory $\boldsymbol{r}_1^*$, the target
point $\boldsymbol{r}_С$ and a~couple of trajectories from a~family
(dashed lines) for the case $\mu _1 =0.2$, $\mu _2 =0.5$. The shaded area is
a~domain of attraction of the system.} \label{fig2} \hfill
\end{figure}

Set the initial segment $[AB]\in Ox$, the initial values of velocities of
the ball $\boldsymbol{v}_0 =5\boldsymbol{j}$,
$\boldsymbol{e}=\frac{1}{2}\boldsymbol{i}+\frac{\sqrt 3 }{2}\boldsymbol
{j}$, the coefficient of friction of the surface in the first phase of
motion $\mu _1 =0.2$ and suppose that the supporting trajectory emanates
from the point $\boldsymbol{r_0}^*=\boldsymbol{i}$. Construct the
trajectory and calculate the corresponding time $t$ of the motion of the
ball until complete stop assuming that the entire surface is homogeneous
with $\mu_1=0.2$. Construct a~supporting curve according to~\eqref{eq5}.
Choose on the trajectory the point $\boldsymbol{r}_1^*(t_1<t)$ at which
the value of the friction coefficient changes from $\mu_1$ to $\mu_2$.
Define on the trajectory of the second phase of motion the target point
$\boldsymbol{r}_C(s)$ through which all trajectories emanating from the
$\delta$-neighborhood of the point $\boldsymbol{r}_0^*$ must pass.
Construct according to~\eqref{eq19} the boundary curve between two
surfaces and the trajectories of motion of balls sufficiently close to the
supporting one. Рассмотрим два случая:

\begin{enumerate}
  \item Set $\mu _2 =0.1$, i.e. the ball passing through the boundary
      between the two surfaces reaches the smoother surface. For this
      case the supporting trajectory, the target point, boundary curve
      between two surfaces, the trajectories of the family and the
      domain of attraction are shown in Fig.~\ref{fig1}. The curve of
      the boundary between the two surfaces has the form of a~convex
      lens, and  the $\delta$-neighborhood (domain of attraction)
      which the trajectories leave before converging to a~target point
      is about $0.007$ meters.

  \item Set $\mu _2 =0.5$, i.e. the ball passing through the boundary
      between the two surfaces reaches the rougher surface. The
      supporting curve, the target point, the boundary curve between
      two surfaces, the trajectories of the family and the domain of
      attraction are shown in Fig.~\ref{fig2}. The curve of the
      boundary between two surfaces has the form of a~slightly concave
      lens, and  the $\delta$-neighborhood (domain of attraction)
      which the trajectories leave before converging to a~target point
      is about $0.001$ meters.
\end{enumerate}

\textbf{The case of bowling balls}

\begin{figure} [!ht]
\centering
\includegraphics{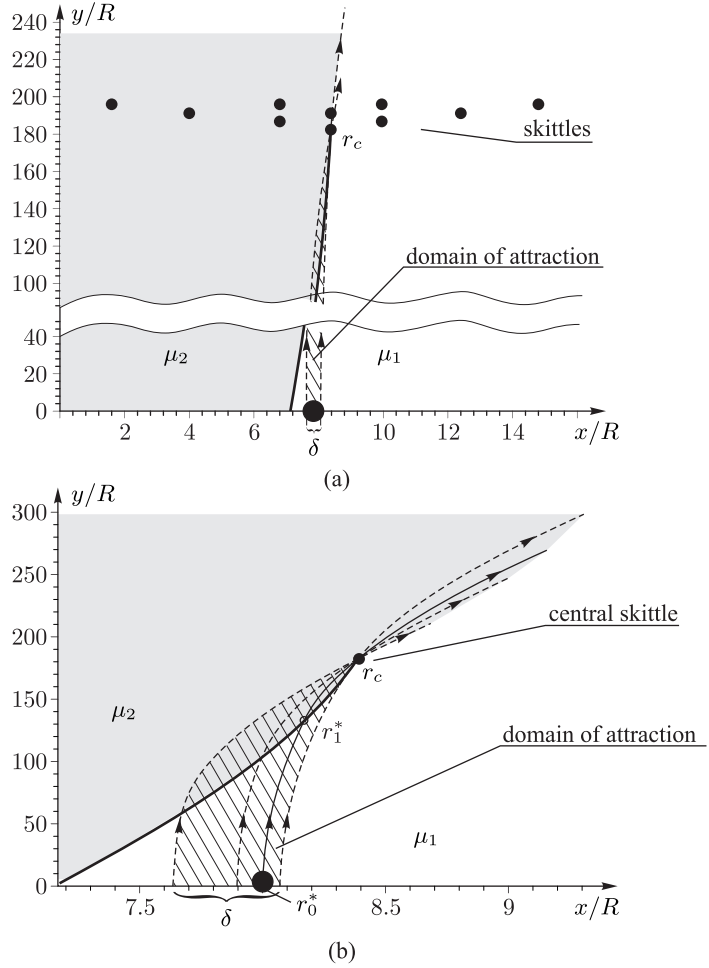}
\caption{ The system similar to the system of a~sliding ball in bowling in
two different scales: (a) the scale that is closer to the real bowling
system, (b)~the scale with an increased and more detailed domain of
attraction. The thin solid line is the supporting trajectory, the thick
solid line is the curve of the boundary between two surfaces is ,
$\boldsymbol{r}_1^*$ is the point of the boundary between two surfaces on
the supporting trajectory and $\boldsymbol{r}_С$ is the target point. The
dashed lines are the trajectories from a~family, the shaded area is a
domain of attraction of the system. The coefficients of friction are $\mu
_1 =0.04$, $\mu _2 =0.2$, and the initial velocities are
$\boldsymbol{v_0}=[0;10]$~m/s, $\boldsymbol{\omega_0}=[0;5]$~s$^{-1}$.}
\label{fig3}
\end{figure}

Now consider the dynamics of a~system similar to the system of a~sliding
ball in bowling. Set $m=1$~kg, $R=0.1$~m, $\mu_1=0.04$, $\mu_2=0.2$,
$\boldsymbol{v_0}=[0;10]$~m/s, $\boldsymbol{\omega_0}=[0;5]$~s$^{-1}$,
$\boldsymbol{r_0}^*=0.8$~m. We can calculate the values $u_0\approx 10$ m/s
and $\boldsymbol{e}\approx 0.05\boldsymbol{i}+0.1\boldsymbol{j}$. Let us
construct a~supporting trajectory and choose the time of motion on
different surfaces so that the target point $\boldsymbol{r_C}$ coincides
with the position of the central skittle. Further, let us construct the
curve of boundary between two surfaces and the trajectories of motion of
the balls leaving the $\delta$-neighborhood (domain of attraction) and
converging to the target point (see Fig.~\ref{fig3}). The calculations
have shown that the boundary between the two surfaces has the form of a
slightly concave lens as in the previous case. The $\delta$-neighborhood
which the trajectories leave before converging to the target point is
about $0.05$ meters. The other trajectories from the $\delta$-neighborhood
don't reach the target point since the sliding motion terminates earlier.

\section{Discussion}

Assuming the coefficient of friction to be variable, we have presented the
algorithm and constructed the curves of the boundary between two surfaces
on a~plane in such a~way that parallel families of analogous homogenous
sliding balls launched under equal initial conditions (linear and angular
velocities) on a~horizontal rough plane, converges to a~predetermined
point. Calculations are presented for some arbitrary cases of passage to
the smoother or to the rougher surface, and also for a~system similar to
that of a~sliding ball in bowling before the transition to rolling.

Numerical experiments have shown that the relations between the s of the
initial positions of the balls in the families and dimensions of a~sliding
domain are very small, and the curves of the boundary between two surfaces
for the cases of passage to the smoother or to the rougher surface  have
qualitative differences.

For example, for the family of balls passing during sliding to the
smoother surface (see Fig.~\ref{fig1}), the curve of the boundary between
two surfaces leaves the abscissa axis on the right of the supporting
trajectory and reaches the target point above the supporting trajectory
and has the form of a~convex lens. For the family passing during sliding
to the rougher surface (see Fig.~\ref{fig2}), the curve of the boundary
between two surfaces leaves the abscissa axis on the left of the
supporting trajectory and reaches the target point below the supporting
trajectory and has the form of a~slightly concave lens.

As for possible application of this effect of refraction of trajectories
for the bowling game, it seems that a~novice player has to train hard at
first to inscribe the ball into a~small $\delta$-neighborhood of the
starting point of the supporting trajectory and to impart the required
initial velocities to  the ball. Also, it is clear that the dimensions of
the domain of attraction ($\delta$-neighborhood) are rather small, about
$5$~cm (see Fig.~\ref{fig3}).

It should be noted that the model considered  can be applicable under
quite rare conditions of the ball's motion~--- pure sliding without
rolling. Incorporating the rolling motion adds the realism to the system,
but implies consideration of more complicated model, for example, the
nonholonomic model extensively studied in~\cite{BMK4}, particular
motion in the limiting case of passage of the ball from an absolutely
rough to an absolutely smooth surface is considered in~\cite{cortes}.

\section{Appendix. Estimate of possible dimensions of domain of attraction}

One way to enlarge the domain of attraction is to increase the number of
the curves of a~boundary between surfaces $\varphi (r)$ in~(\ref{eq8}).
Physically this can be done by gluing figured strips from materials with
different coefficients of friction onto the floor. The form of these
strips can be calculated by analogy with Sections~\ref{sec2}--\ref{sec5}.
It should be kept in mind that, in contrast to the case of a~single
switching, the form of the strips is not uniquely defined by the
condition~\eqref{eq9}, and it is necessary to add some optimization
requirement, which complicates the problem significantly.

An estimate of the maximum of the domain of attraction can be found by
considering the family of parabolas~\eqref{eq7} and~\eqref{eq8} without
switching. Each of the parabolas is characterized by its coefficient of
friction $f(\delta )$, which corresponds to a~continuous change of the
coefficient of friction on the supporting plane. The boundary values for
the initial conditions $\left( {\delta _{\min } ,\delta _{\max } }
\right)$ correspond to the limiting values
\[
f(\delta _{\min } )=f_{\max } ,\quad f(\delta _{\max } )=0,
\]
where the value $f_{\max } $ corresponds to the maximally rough material
used. We define the target point $C$ as the intersection of the limit
trajectories
\begin{equation}
\label{eq101}
\boldsymbol {r}_0^\ast +\delta _{\max } \boldsymbol {i}+\boldsymbol {v}_0 t_1 =\boldsymbol {r}_C ,\quad
\boldsymbol {r}_0^\ast +\delta _{\min } \boldsymbol {i}+\boldsymbol {v}_0 t_2 +\frac{1}{2}f_{\max
} t_2^2 \boldsymbol {e}=\boldsymbol {r}_C,
\end{equation}
where $\delta _{\min } $ and $t_2 $ are given arbitrarily, and $\delta
_{\max } $ and $t_1 $ are found from the vector equality
\begin{equation}
\label{eq102}
\left( {\delta _{\max } -\delta _{\min } } \right)(\alpha \boldsymbol {v}_0 +\beta
\boldsymbol {e})+\boldsymbol {v}_0 t_1 =\boldsymbol {v}_0 t_2 +\frac{1}{2}f_{\max } t_2^2 \boldsymbol
{e}.
\end{equation}
For the values $\delta \in \left( {\delta _{\min } ,\delta _{\max } }
\right)$ the quantity $f(\delta )$ is chosen such that the
parabola~\eqref{eq8} with this coefficient crosses the target point. Thus,
the lines of the level set of the function $f$ on the supporting plane
have the form of parabolas (in the limit $f\to 0$ is a~straight line).

Equating the coefficients with $\boldsymbol {e}$ in~\eqref{eq102}, we
obtain
\begin{equation}
\label{eq103}
\left( {\delta _{\max } -\delta _{\min } } \right)\beta =\frac{1}{2}\mu_{\max
}g t_2^2.
\end{equation}

Formula~\eqref{eq103} shows that the dimension of the domain of attraction
is proportional to the maximum coefficient of friction and to the square
of time of the motion of a~ball into the target point. If we set
$\mu=0.3$, $t=4$ seconds, we obtain 24 meters of the maximum possible
length of the domain of attraction.

\section{Acknowledgements}

The authors are grateful to Oliver O'Reilly, Alexey Borisov, Ivan Mamaev and Tatiana Ivanova for useful discussions
and valuable remarks.  The N. Erdakova's work was supported by the Grant RFBR №15-08-09261-а.
The A. Ivanov's work was supported by the RFBR grant
14-01-00432 and was carried out within the framework of the basic part of the state assignment №
2014/120.

\end{document}